\documentclass[a4paper,11pt]{article}
\usepackage{cite}
\usepackage{float}
\usepackage{draft}
\usepackage{mathtools}
\usepackage{bm}
\usepackage{booktabs}
\usepackage{array}
\usepackage{tabularx}
\usepackage{microtype}
\usepackage{enumitem}
\usepackage{amsthm}
\usepackage{orcidlink}
\usepackage[nameinlink,capitalise,noabbrev]{cleveref}
\usepackage{tikz}
\usetikzlibrary{arrows.meta,positioning,shadows.blur}
\usetikzlibrary{arrows.meta,positioning,fit}
\allowdisplaybreaks[4]
\definecolor{adhmblue}{RGB}{39,92,150}
\definecolor{abjmred}{RGB}{162,57,65}
\definecolor{bmngreen}{RGB}{40,118,91}
\definecolor{softgray}{RGB}{246,247,249}
\definecolor{gold}{RGB}{181,130,36}
\definecolor{linkblue}{RGB}{25,70,120}
\definecolor{citegreen}{RGB}{35,105,80}
\definecolor{urlteal}{RGB}{20,105,115}
\hypersetup{
  breaklinks=true,
  colorlinks=true,
  linkcolor=linkblue,
  citecolor=citegreen,
  urlcolor=urlteal,
  pdftitle={Equal-GNO Sectors and Fuzzy-Sphere Vacua: Large-Charge ADHM--BMN Index Matching},
  pdfauthor={Sarthak Duary; Kangning Liu},
  pdfsubject={Fixed-sector matching between the ADHM superconformal index and the BMN matrix-model index},
  pdfkeywords={ADHM theory, BMN matrix model, superconformal index, GNO charge, fuzzy sphere}
}

\crefformat{section}{section~#2#1#3}
\crefformat{subsection}{section~#2#1#3}
\crefformat{subsubsection}{section~#2#1#3}
\crefrangeformat{section}{sections~#3#1#4--#5#2#6}

\newcommand{\ADHM}{\mathrm{ADHM}}
\newcommand{\BMN}{\mathrm{BMN}}
\newcommand{\Tr}{\operatorname{Tr}}
\newcommand{\cQ}{\mathcal Q}

\newcommand{\Mcal}{\mathcal M}
\newcommand{\dd}{\mathrm d}
\newcommand{\ii}{\mathrm i}
\newcommand{\cI}{\mathcal I}
\newcommand{\cK}{\mathcal K}
\newcommand{\cO}{\mathcal O}
\newcommand{\bZ}{\mathbb Z}
\newcommand{\Poch}[2]{(#1;#2)_\infty}
\newcommand{\PochF}[3]{(#1;#2)_{#3}}
\newcommand{\one}{\mathbf 1}
\newcommand{\cZ}{\mathcal Z}

\title{Large-charge ADHM--BMN index matching}

\affiliation[a]{Yau Mathematical Sciences Center (YMSC), Tsinghua University,
Beijing, China}
\affiliation[b]{Department of Mathematical Sciences, Tsinghua University,
Beijing, China}
\affiliation[c]{Jefferson Physical Laboratory, Harvard University, Cambridge, Massachusetts 02138, USA}

\author[a,\orcidlink{0000-0002-4535-3198}]{Sarthak Duary}
\emailAdd{sarthakduary@tsinghua.edu.cn}
\author[a,b,c,\orcidlink{0009-0009-5266-3600}]{and Kangning Liu} \emailAdd{lkn22@mails.tsinghua.edu.cn}

\abstract{We compare the superconformal index of three-dimensional ADHM
theory in fixed monopole sectors with the index of the BMN matrix model
around fuzzy-sphere vacua. The monopole charges determine the sizes of
the fuzzy-sphere blocks, and their multiplicities determine how many
identical blocks occur. At a fixed number of fundamental hypermultiplets,
we increase all positive monopole charges by the same amount while keeping
their differences and multiplicities fixed. After dividing the ADHM index
by the contribution of the bare monopole, it agrees order by order with
the BMN index in the corresponding limit of large blocks. Fundamental
excitations move to arbitrarily high orders in the index expansion, while
the upper angular-momentum cutoffs of the BMN harmonics diverge, leaving
the same vector and adjoint
contribution. }

\begin{document}
\maketitle

\section{Introduction}
\label{sec:introduction}

Supersymmetric indices isolate short representations while discarding the
paired long-multiplet spectrum. Their robustness allows a protected
quantity defined at strong coupling to be computed by localization or by a
weakly coupled oscillator description \cite{Witten:1982df,Kinney:2005ej}.
In three dimensions the superconformal index on $S^2\times S^1$ has an
additional structural feature: it is a sum over monopole sectors labelled
by Goddard--Nuyts--Olive (GNO) monopole charges
\cite{Goddard:1976qe,Borokhov:2002ib}. A monopole sector fixes the full
charge vector up to Weyl transformations, which act as permutations for
$U(N)$. Each term is a contour integral over holonomies, weighted by
one-loop determinants in a prescribed monopole background. The
localization formula and its applications to three-dimensional
superconformal theories were developed in
\cite{Bhattacharya:2008zy,Bhattacharya:2008bja,Kim:2009wb,
Imamura:2011su,Yokoyama:2011wi,Kapustin:2011jm,Krattenthaler:2011da}.

The BMN matrix model is a massive
deformation of BFSS matrix quantum mechanics
\cite{Banks:1996vh,Taylor:2001vb,Berenstein:2002jq}. It furnishes a
discrete-light-cone description of M-theory in the maximally supersymmetric
eleven-dimensional plane wave. The mass deformation lifts the flat
directions of BFSS and replaces the continuous classical moduli space by a
set of isolated supersymmetric vacua. These vacua are fuzzy two-spheres: the three distinguished matrices form a generally reducible $SU(2)$ representation which can be decomposed into irreducible representations, so every vacuum is labelled by a partition of the matrix size. The multiplicity of an irreducible block is the rank of the unbroken unitary group acting on identical blocks. Perturbation theory and the protected spectrum around these vacua were analysed in
\cite{Dasgupta:2002hx,Kim:2002wg,Dasgupta:2002ru,Kim:2002qz,
Maldacena:2002rb,Lin:2005nh}. More recently, the refined BMN Witten index
was written as a unitary-matrix integral in an arbitrary vacuum sector
\cite{Chang:2024zwt}, and the finite-matrix-size sum over all sectors was studied in \cite{Chang:2026allsectors}.

In this work we compare the BMN index in a fixed vacuum with the
three-dimensional ADHM superconformal index in a specified monopole
sector. The motivation comes from IR dualities
among three-dimensional theories describing M2-branes.
Figure~\ref{fig:SYM-ADHM-ABJM} shows their UV descriptions:
SYM theory on D2-branes, ADHM
theory, and ABJM theory. Here $N_f$ counts fundamental hypermultiplets
in ADHM theory, and $k$ is the ABJM Chern--Simons level.
\begin{figure}[H]
  \centering
  \begin{tikzpicture}[
    theory/.style={
      minimum width=3.75cm,
      minimum height=2.25cm,
      align=center,
      inner sep=7pt
    },
    duality/.style={
      <->,
      >=Latex,
      very thick,
      shorten <=3pt,
      shorten >=3pt
    },
    arrowlabel/.style={
      align=center,
      font=\scriptsize,
      fill=white,
      inner xsep=1.2pt,
      inner ysep=0.5pt,
      outer sep=0pt
    },
    node distance=1.9cm
  ]

  \node[
    theory
  ] (SYM) {
    \textbf{\(\boldsymbol{\mathcal N=8}\) SYM}\\[2mm]
    \(U(N)\) gauge theory\\[1mm]
    \footnotesize D2-brane description
  };

  \node[
    theory,
    right=of SYM
  ] (ADHM) {
    \textbf{\(\boldsymbol{N_f=1}\) ADHM}\\[2mm]
    \(U(N)\) gauge theory\\[1mm]
    \footnotesize
    UV: \(\mathcal N=4\),\quad
    IR: \(\mathcal N=8\)
  };

  \node[
    theory,
    right=of ADHM
  ] (ABJM) {
    \textbf{\(\boldsymbol{k=1}\) ABJM}\\[2mm]
    \(U(N)_1\times U(N)_{-1}\)\\[1mm]
    \footnotesize
    \(N\) M2-branes on \(\mathbb C^4\)
  };

  \draw[duality]
    (SYM.east) --
    node[
      arrowlabel,
      midway,
      above=2.2mm
    ] {\(\textit{mirror}/\textit{IR}\)}
    (ADHM.west);

  \draw[duality]
    (ADHM.east) --
    node[
      arrowlabel,
      midway,
      above=2.2mm
    ] {\(\textit{IR duality}\)}
    (ABJM.west);

  \node[
    below=8mm of ADHM,
    align=center,
    font=\small\itshape,
    text width=8cm
  ] {
    Proposed infrared descriptions of the
    \(\mathcal N=8\) M2-brane fixed point
  };

  \end{tikzpicture}

  \caption{
    The proposed infrared duality network relating maximally supersymmetric
    \(U(N)\) Yang--Mills theory, the \(N_f=1\) ADHM theory, and the
    \(k=1\) ABJM theory. Although their ultraviolet Lagrangian
    descriptions are different, they are expected to flow to the same
    \(\mathcal N=8\) M2-brane fixed point.
  }
  \label{fig:SYM-ADHM-ABJM}
\end{figure}

For $N_f=1$, the ADHM theory has enhanced $\mathcal N=8$
supersymmetry in the infrared and is conjecturally equivalent to
$U(N)_1\times U(N)_{-1}$ ABJM theory. Agreement of sphere partition
functions and refined superconformal indices supports this equivalence
\cite{Kapustin:2010xq,Hayashi:2022ldo}. This relation motivates an
M2-brane interpretation of the comparison. The index limit derived below
holds for any fixed $N_f\geq1$.

The $U(N)$ ADHM theory contains one adjoint hypermultiplet and
$N_f$ fundamental hypermultiplets and is engineered on D2-branes probing
D6-branes \cite{Douglas:1995bn,Hayashi:2022ldo}. Its superconformal index
is a sum over integer GNO charge vectors $\bm m=(m_1,\ldots,m_N)$,
with the flux normalization specified in \eqref{eq:coulomb-monopole-saddle}.
We use the supersymmetric monopole saddles of Coulomb-branch localization on
$S^2\times S^1$: the magnetic flux is accompanied by the real
vector-multiplet scalar, and the hypermultiplet scalar backgrounds
vanish \cite{Imamura:2011su}. The saddle and its relation to monopole
operators are reviewed in \cref{sec:coulomb-saddle}.

We pair a monopole sector with $s$ distinct positive charges,
represented by
\begin{equation}
  \bm m=
  \bigl(
  \underbrace{N_1,\ldots,N_1}_{n_1},
  \ldots,
  \underbrace{N_s,\ldots,N_s}_{n_s}
  \bigr),
  \qquad N_\alpha>0,
  \label{eq:intro-equal-sector2}
\end{equation}
with a BMN vacuum containing $n_\alpha$ blocks of dimension $N_\alpha$.
The ADHM rank is $N=\sum_\alpha n_\alpha$, the BMN matrix size is
$N_{\mathrm{BMN}}=\sum_\alpha n_\alpha N_\alpha$, and the residual
gauge group on both sides is $H=\prod_\alpha U(n_\alpha)$, where
$\alpha$ labels the distinct monopole charges. We apply a common charge
shift
\begin{equation}
  N_\alpha=\Lambda+\nu_\alpha,
  \qquad \Lambda\longrightarrow\infty,
  \qquad N_\alpha-N_\beta=\nu_\alpha-\nu_\beta,
  \label{eq:intro-common-shift}
\end{equation}
at fixed integers $\nu_\alpha$, multiplicities $n_\alpha$ and
number of flavors $N_f$.

On the ADHM side, the vector and adjoint determinants depend only on
the fixed differences $\nu_\alpha-\nu_\beta$. Fundamental fields carry
the absolute charges $N_\alpha$, so their first oscillator contributions
move to arbitrarily high orders in $q$. The fugacities $q$, $\xi_F$ and
$\xi_T$ are defined in the index trace \eqref{eq:ADHM-trace} below;
$\xi_F$ and $\xi_T$ couple to the flavor and topological charges,
respectively. The BMN fugacity $c=q\xi_F^{-1}$ is related to the ADHM
fugacities by \eqref{eq:adhm-to-bmn-map}. The bare monopole contributes
$\xi_T^P c^{N_fP/2}$, where $P=\sum_\alpha n_\alpha N_\alpha$ is the
total topological charge. After dividing the sector contribution by this factor,
the remaining fundamental determinant tends to one order by order in
$q$. On the BMN side, the minimum angular momentum of the rectangular
fuzzy-spherical harmonics is $|N_\alpha-N_\beta|/2$, which remains
fixed, while their upper cutoffs diverge. The resulting infinite towers
give the same vector--adjoint kernel as the ADHM monopole harmonics.

We prove the equality of the two limiting matrix integrals in
\eqref{eq:general-matching}. At finite charge the two indices have
different corrections: fundamental excitations on the ADHM side and
missing high-spin harmonics on the BMN side. Their orders are bounded
in \eqref{eq:general-integrated-bounds}.

\Cref{sec:physical-dictionary} reviews the D2--D6 brane construction
and relates ADHM monopole backgrounds to BMN fuzzy-sphere vacua.
\Cref{sec:indices-review} reviews the BMN and ADHM indices and fixes
the relation between their charges and fugacities.
\Cref{sec:matching} first derives the matching for equal charges, then
extends the calculation to general positive charges and establishes the
common large-charge limit. It also bounds the finite-charge corrections
and checks an example with unequal charges.
\Cref{sec:conclusions} summarizes the results and discusses operator maps
to ABJM and mirror SYM, together with further finite-charge calculations.

\textit{Note Added.} During the final preparation of this work, we learned of the forthcoming work \cite{Komatsu:2026part2}, which is mentioned in the Discussion section of Ref.~\cite{Komatsu:2026ano}. We note that the limit considered there is different from the limit studied in this work. 

\section{Physical setting}
\label{sec:physical-dictionary}
The D2--D6 construction identifies ADHM topological charge with momentum
along the M-theory circle. The monopole charges determine the scalar
background, the mass gap of the fundamental hypermultiplets and the
sizes of the corresponding BMN blocks.

\subsection{Brane construction}
\label{sec:d2d6}
\label{sec:circle-momentum}
\label{sec:mirror-sym}

The ADHM theory is the three-dimensional $\mathcal N=4$ $U(N)$ gauge
theory on $N$ D2-branes probing $N_f$ D6-branes. D2--D2 strings supply the vector multiplet and one adjoint
hypermultiplet, and D2--D6 strings supply $N_f$ fundamental
hypermultiplets. Their $\mathcal N=2$ decomposition is listed in
\cref{tab:adhm-fields}.

\begin{table}[H]
\centering
\renewcommand{\arraystretch}{1.18}
\begin{tabular}{@{}lll@{}}
\toprule
\(\mathcal N=4\) multiplet & \(\mathcal N=2\) fields & \(U(N)\) representation\\
\midrule
Vector multiplet
& vector \(V\) and chiral \(\Phi_3\)
& adjoint\\
Adjoint hypermultiplet
& chirals \(\Phi_1,\Phi_2\)
& adjoint\\
\(N_f\) fundamental hypers
& \(\psi_q,\widetilde\psi_q\), \(q=1,\ldots,N_f\)
& fundamental and anti-fundamental\\
\bottomrule
\end{tabular}
\caption{Multiplet content of the three-dimensional ADHM model.}
\label{tab:adhm-fields}
\end{table}

The \(\mathcal N=4\) superpotential, with its overall normalization
suppressed, is (see equation~(25) of \cite{Bashkirov:2010kz} for
$N_f=1$)
\begin{equation}
W
=
\Tr\!\left[
\Phi_3\left([\Phi_1,\Phi_2]+\sum_{q=1}^{N_f}\psi_q\widetilde\psi_q\right)
\right].
\label{eq:adhm-superpotential}
\end{equation}
A convenient brane orientation is
\begin{equation}
\begin{array}{c|cccccccccc}
 &0&1&2&3&4&5&6&7&8&9\\ \hline
\mathrm{D2}&\times&\times&\times&-&-&-&-&-&-&-\\
\mathrm{D6}&\times&\times&\times&\times&\times&\times&\times&-&-&-
\end{array}.
\label{eq:brane-array}
\end{equation}
D2-brane motion along directions $3,4,5,6$, which lie within the
D6-branes, gives the adjoint-hypermultiplet scalars. Motion along the
common transverse directions $7,8,9$ gives the vector-multiplet scalar
triplet $(\sigma,\operatorname{Re}\Phi_3,\operatorname{Im}\Phi_3)$.
Magnetic flux on the D2 worldvolume carries dissolved Dirichlet zero-brane
(D0-brane) charge through
the Wess--Zumino coupling \cite{Douglas:1995bn}; equal diagonal flux
distributes this charge uniformly among the D2-branes.

Upon lifting type IIA string theory to M-theory, the D2-branes become
M2-branes and each D6-brane becomes a Kaluza--Klein monopole (KK6),
\begin{equation}
  \mathrm{D2}\longrightarrow\mathrm{M2},
  \qquad
  \mathrm{D6}\longrightarrow\mathrm{KK6}.
  \label{eq:m-lift}
\end{equation}
Near a stack of \(N_f\) coincident D6-branes, the transverse geometry becomes the
\(A_{N_f-1}\) singularity
\begin{equation}
\mathbb C^2/\mathbb Z_{N_f}.
\label{eq:orbifold}
\end{equation}
Including the adjoint-hypermultiplet directions gives the infrared
geometry
\begin{equation}
\mathbb C^2\times\frac{\mathbb C^2}{\mathbb Z_{N_f}}.
\label{eq:m2-orbifold}
\end{equation}
The transverse geometry of one D6-brane is Taub--NUT space. Its fibre is
the M-theory circle, and its one-centre core is locally
$\mathbb R^4\simeq\mathbb C^2$. For $N_f=1$, this Taub--NUT
$\mathbb C^2$ combines with the $\mathbb C^2$ parametrized by the adjoint
hypermultiplet to give the eight real transverse directions of an M2-brane
\cite{Aharony:2008ug,Hayashi:2022ldo}.

\begin{figure}[H]
\centering
\begin{tikzpicture}[
  node distance=11mm and 11mm,
  theory/.style={align=center, minimum height=12mm, text width=39mm},
  arrow/.style={-{Latex[length=2.4mm]}, thick},
  note/.style={align=center, font=\small, text width=34mm}
]
\node[theory] (branes)
{\textbf{Type IIA branes}\\
\(N\) D2 + \(N_f\) D6};
\node[theory, right=of branes] (gauge)
{\textbf{\(\mathcal N=4\) ADHM theory}\\
\(U(N)\), adjoint hypermultiplet, \(N_f\) fundamentals};
\node[theory, right=of gauge] (mtheory)
{\textbf{M-theory lift}\\
\(N\) M2 on
\(\mathbb C^2\times\mathbb C^2/\mathbb Z_{N_f}\)};
\draw[arrow, adhmblue] (branes) -- (gauge);
\draw[arrow, bmngreen] (gauge) -- (mtheory);
\node[note, below=7mm of branes] {D2--D2 strings give\\matrix fields};
\node[note, below=7mm of gauge] {D2--D6 strings give\\fundamental matter};
\node[note, below=7mm of mtheory] {D6 becomes an\\\(A_{N_f-1}\) singularity};
\end{tikzpicture}
\caption{The D2--D6 construction and its M-theory lift. The number $N_f$
of fundamental hypermultiplets equals the number of D6-branes and the
orbifold order.}
\label{fig:adhm-origin}
\end{figure}
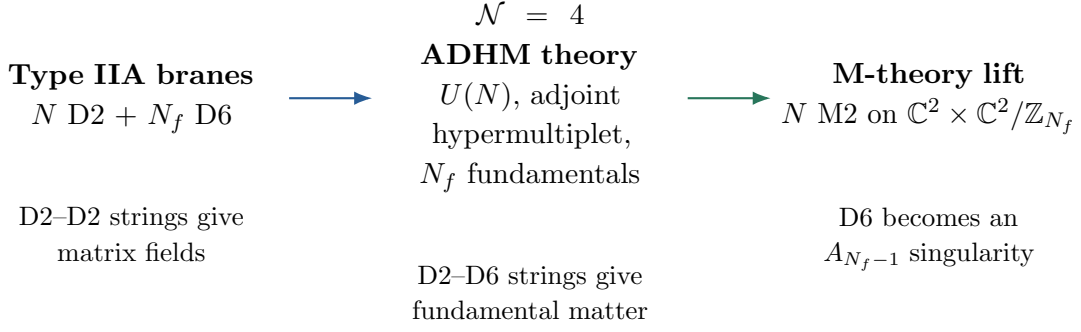

The central $U(1)$ gauge field in three dimensions can be dualized to a
periodic scalar. In the D2--D6 lift this dual photon is the angular
coordinate of the Taub--NUT fibre. With
$F=\tfrac12F_{\mu\nu}\,\dd x^\mu\wedge\dd x^\nu$, its conserved
topological current and charge are
\begin{equation}
  j_T^\mu=\frac{1}{2\pi}(*\Tr F)^\mu
       =\frac{1}{4\pi}\epsilon^{\mu\nu\rho}\Tr F_{\nu\rho},
  \qquad
  Q_T=\frac{1}{2\pi}\int_{S^2}\Tr F
      =\sum_{i=1}^{N}m_i .
  \label{eq:topological-charge}
\end{equation}
A topological rotation translates the dual photon, so $Q_T$ is the integer
momentum conjugate to the M-theory-circle angle. We choose the orientation
in which positive GNO charge gives positive circle momentum. In a fixed
sector, the topological fugacity contributes the overall factor
$\xi_T^{Q_T}$ associated with the fixed longitudinal momentum.

For $N_f=1$, the ADHM theory and three-dimensional $\mathcal N=8$
$U(N)$ super Yang--Mills flow, with the mirror choice of R-symmetry,
to the same M2-brane fixed point
\cite{Gang:2011jj,Cremonesi:2013lqa}. In $\mathcal N=4$ language,
the SYM theory contains a vector multiplet and an adjoint hypermultiplet.
Mirror symmetry maps the ADHM topological symmetry to an ordinary
flavor symmetry in this frame. A fixed total ADHM topological charge
$T=Q_T$ therefore selects a flavor-charge sector in mirror SYM.

\subsection{Matching the vacua}
\label{sec:coulomb-saddle}
\label{sec:charges-vacua}
\label{sec:orbit-dictionary}
\label{sec:equal-physics}

We set all real and complex masses and all Fayet--Iliopoulos parameters
of the ADHM theory to zero. Semiclassically, vacua on the Coulomb branch
have vanishing hypermultiplet scalars and commuting vector-multiplet
scalars. In our notation, the latter are $\sigma$,
$\operatorname{Re}\Phi_3$ and $\operatorname{Im}\Phi_3$.
A classical vacuum has constant scalar expectation values obeying
\begin{equation}
 \Phi_1=\Phi_2=\psi_q=\widetilde\psi_q=0,\qquad
 [\sigma,\Phi_3]=[\Phi_3,\Phi_3^\dagger]=0,
 \qquad \sigma=\sigma^\dagger.
 \label{eq:adhm-coulomb-directions}
\end{equation}
At a generic point with distinct vector-multiplet scalar eigenvalues,
the gauge group is broken to $U(1)^N$. Dualizing the abelian gauge fields
gives $N$ periodic dual photons, so the Coulomb branch has real dimension
$4N$. In the complex structure selected by our $\mathcal N=2$
description, the chiral ring of the quantum Coulomb branch is generated
by gauge-invariant polynomials in $\Phi_3$ and
BPS monopole operators,
including dressed monopole operators
\cite{Bullimore:2015lsa,Cremonesi:2013lqa}.

For the massless ADHM theory, the quantum Coulomb branch, viewed as
a complex algebraic variety, is
\begin{equation}
 \Mcal_{\mathrm C}
 =\operatorname{Sym}^{N}\!\left(\mathbb C^2/\mathbb Z_{N_f}\right),
 \qquad
 \left.\Mcal_{\mathrm C}\right|_{N_f=1}
 =\operatorname{Sym}^{N}(\mathbb C^2),
 \label{eq:adhm-coulomb-variety}
\end{equation}
as obtained in section~4.1 of \cite{Cremonesi:2013lqa}. Here
$\operatorname{Sym}^{N}$ denotes the symmetric-product affine variety.
For the BPS monopole operators and the
Coulomb-branch chiral-ring relations of the ADHM theory, see
sections~2 and~5.3.2 of \cite{Bullimore:2015lsa}.

For Coulomb-branch localization of the $S^2\times S^1$ superconformal
index, let $R_{S^2}$ be the sphere radius and use orthonormal indices
$1,2$ on the sphere and $3$ along Euclidean time. In the conventions of \cite{Imamura:2011su}, the bosonic vector-multiplet
localizing term is
\begin{equation}
 S^{\mathrm{vec}}_{\mathrm{loc,bos}}
 =t\int_{S^2\times S^1}\dd^3x\sqrt{g}\,
 \Tr\!\left[\sum_{\mu=1}^{3}\mathcal V_\mu^2+D_{\mathrm{aux}}^2\right],
 \qquad t>0,
 \label{eq:coulomb-localizing-action}
\end{equation}
where
\begin{equation}
 \begin{aligned}
 \mathcal V_1&=F_{23}-D_1\sigma,&
 \mathcal V_2&=F_{31}-D_2\sigma,\\
 \mathcal V_3&=F_{12}-D_3\sigma-\frac{\sigma}{R_{S^2}}.
 \end{aligned}
 \label{eq:coulomb-localizing-equations}
\end{equation}
The symbol $D_{\mathrm{aux}}$ denotes the real auxiliary field. On the localization contour, the squares vanish
when $\mathcal V_\mu=D_{\mathrm{aux}}=0$.

For a static, purely magnetic background, impose
$F_{3a}=D_3\sigma=0$ for $a=1,2$. Substitution into
\eqref{eq:coulomb-localizing-equations} gives
\begin{equation}
 F_{3a}=0,\qquad D_3\sigma=D_a\sigma=0,\qquad
 F_{12}=\frac{\sigma}{R_{S^2}},\qquad D_{\mathrm{aux}}=0.
 \label{eq:coulomb-static-bps}
\end{equation}
We take the scalar backgrounds of all $\mathcal N=2$ chiral multiplets to
vanish:
\begin{equation}
 \Phi_1=\Phi_2=\Phi_3=\psi_q=\widetilde\psi_q=0.
 \label{eq:coulomb-zero-chiral-backgrounds}
\end{equation}
The fermionic and chiral auxiliary backgrounds also vanish, as do all
derivatives of \eqref{eq:adhm-superpotential}. The localization
determinants retain the fluctuations of every chiral multiplet about
these backgrounds. These include $\Phi_3$ dressings of the bare monopole,
formed from polynomials invariant under the subgroup preserving its
monopole charge \cite{Bullimore:2015lsa}.

Using $F_{\theta\phi}=R_{S^2}^2\sin\theta\,F_{12}$, integrating
\eqref{eq:coulomb-static-bps} over the sphere and imposing flux
quantization gives
\begin{equation}
  {\begin{aligned}
 \frac{1}{2\pi}\int_{S^2}F
   &=\operatorname{diag}(m_1,\ldots,m_N),\\
 \sigma&=\frac{\operatorname{diag}(m_1,\ldots,m_N)}{2R_{S^2}},\\
 F_{\theta\phi}
   &=\frac{\operatorname{diag}(m_1,\ldots,m_N)}2\sin\theta .
 \end{aligned}}
 \label{eq:coulomb-monopole-saddle}
\end{equation}
Here $m_i\in\mathbb Z$, with permutations identified. A commuting flat
connection along $S^1$ supplies the holonomy eigenvalues
$z_i=e^{\ii h_i}$, where $h_i$ are the eigenvalues of
$\oint_{S^1}A_3\dd x^3$. We retain the integer fluxes $m_i$
throughout. Equations~\eqref{eq:coulomb-static-bps}--\eqref{eq:coulomb-monopole-saddle}
fix our relative sign between flux and $\sigma$. Under radial quantization, a BPS monopole operator creates a state on
$S^2$ with the magnetic flux and scalar background in
\eqref{eq:coulomb-monopole-saddle}
\cite{Borokhov:2002cg,Cremonesi:2013lqa}.

For the bare monopole, the vector and adjoint-hypermultiplet vacuum
contributions cancel. The bare monopole therefore has scaling dimension
\begin{equation}
 \begin{aligned}
 \Delta_0(\bm m)
 &=-\sum_{i<j}|m_i-m_j|
   +\frac12\sum_{i,j}|m_i-m_j|
   +\frac{N_f}{2}\sum_i|m_i|\\
 &=\frac{N_f}{2}\sum_i|m_i|.
 \end{aligned}
 \label{eq:coulomb-bare-dimension}
\end{equation}
The $N_f=1$ calculation and Gauss-law discussion are given in \cite{Bashkirov:2010kz}; the general
$N_f$ result is in \cite{Cremonesi:2013lqa}. At zero
Chern--Simons level, the bare monopole satisfies Gauss law
without charged matter dressing. Its physical cylinder energy is
$\Delta_0/R_{S^2}$.

In the discrete light-cone description of M-theory, the BMN matrix size is
the positive longitudinal-momentum quantum number
\cite{Susskind:1997cw,Sen:1997we,Seiberg:1997ad}. The topological charge
defined in \eqref{eq:topological-charge} therefore sets the total matrix
size. The individual monopole charges specify its decomposition into
irreducible fuzzy-sphere blocks.
Consider a monopole sector represented by the charge vector
\begin{equation}
  \bm m=\bigl(N_1^{\,n_1},N_2^{\,n_2},\ldots,N_s^{\,n_s}\bigr),
  \qquad N_\alpha>0,
  \qquad N=\sum_{\alpha=1}^{s}n_\alpha .
  \label{eq:general-positive-orbit}
\end{equation}
Gauge transformations preserving the diagonal monopole background can
mix only entries with equal charge. The unbroken gauge group is therefore
\begin{equation}
  H_{\ADHM}=\prod_{\alpha=1}^{s}U(n_\alpha).
  \label{eq:adhm-commutant}
\end{equation}

The mass parameter of the BMN model gives a discrete set of
supersymmetric vacua. In these vacua, three matrix coordinates are
proportional to $SU(2)$ generators, while the remaining six vanish
\cite{Berenstein:2002jq,Dasgupta:2002hx}.
A vacuum is therefore specified by a representation of $SU(2)$ on the
matrix indices,
\begin{equation}
  J^i=\bigoplus_{\alpha=1}^{s}
       \left(\one_{n_\alpha}\otimes J^i_{(N_\alpha)}\right),
  \qquad
  N_{\BMN}=\sum_{\alpha=1}^{s}n_\alpha N_\alpha ,
  \label{eq:bmn-partition-dictionary}
\end{equation}
where $J^i_{(N_\alpha)}$ is the $N_\alpha$-dimensional irreducible
$SU(2)$ representation. Each nontrivial irreducible block forms a fuzzy
two-sphere, giving a matrix description of a spherical M2-brane in the
eleven-dimensional plane wave. The block size $N_\alpha$ is the number
of longitudinal momentum units carried by one membrane, while
$n_\alpha$ counts coincident membranes with the same momentum
\cite{Chang:2026yeg}. The partition of $N_{\BMN}$ thus specifies how
the total momentum is distributed among the membranes.

Gauge transformations can mix the $n_\alpha$ identical blocks without
changing the vacuum, giving a $U(n_\alpha)$ factor. The full unbroken
gauge group is
\begin{equation}
  H_{\BMN}=\prod_{\alpha=1}^{s}U(n_\alpha),
  \label{eq:bmn-commutant}
\end{equation}
identical to \eqref{eq:adhm-commutant}. For a single irreducible block,
the unbroken group is $U(1)$. The vacuum consisting of $N_{\BMN}$
one-dimensional blocks has all matrix coordinates equal to zero and
preserves the full $U(N_{\BMN})$ gauge group
\cite{Dasgupta:2002hx}. The correspondence between monopole charges and
BMN blocks is summarized in \cref{tab:dictionary}.

\begin{table}[H]
\centering
\renewcommand{\arraystretch}{1.18}
\begin{tabularx}{\textwidth}{@{}p{0.18\textwidth}XX@{}}
\toprule
Quantity & ADHM monopole sector & BMN vacuum \\
\midrule
$N_\alpha$ & magnitude of a GNO charge
& size of an irreducible fuzzy-sphere block \\
$n_\alpha$ & multiplicity of that charge
& multiplicity of that block \\
$\sum_\alpha n_\alpha$ & three-dimensional gauge rank
& rank of the unbroken gauge group \\
$\sum_\alpha n_\alpha N_\alpha$ & M-circle momentum
& total BMN matrix size \\
$\prod_\alpha U(n_\alpha)$ & unbroken gauge group
& unbroken gauge group \\
\bottomrule
\end{tabularx}
\caption{Correspondence between ADHM monopole sectors and BMN vacua.}
\label{tab:dictionary}
\end{table}

On the ADHM side of this correspondence, the monopole background also
determines the mass gap of the fundamental hypermultiplets through its
scalar profile. Along the transverse direction represented by $\sigma$ in \eqref{eq:brane-array},
the monopole saddle fixes the D2 positions through
\begin{equation}
 \sigma_i=\frac{m_i}{2R_{S^2}},
 \qquad
 \mathsf X_i=2\pi\alpha'\sigma_i,
 \label{eq:brane-monopole-scalar}
\end{equation}
where the sign follows the flux convention in
\eqref{eq:coulomb-static-bps}. The second relation is the usual
identification of the worldvolume scalar with a transverse D2 position.
A D6 displacement $\mathsf X_{6,f}=2\pi\alpha'\mu_f$ in the same
direction gives a real mass $\mu_f$. With the remaining transverse
separations set to zero, the D2--D6 string induces the quadratic coupling
\begin{equation}
 M_{2-6,if}
 =\frac{|\mathsf X_i-\mathsf X_{6,f}|}{2\pi\alpha'}
 =|\sigma_i-\mu_f|,
 \qquad
 \mathcal L_{\mathrm{quad}}\supset M_{2-6,if}^{2}|h_{if}|^2,
 \label{eq:brane-fundamental-mass}
\end{equation}
where $h_{if}$ denotes a fundamental-hypermultiplet scalar. This is the
classical separation-dependent mass of the open-string hypermultiplet
with four mixed Neumann--Dirichlet directions \cite{Polchinski:1996fm}.

We set $\mu_f=0$, as in the index below. The D2--D6 and D2--D2
background masses are then
\begin{equation}
 M_{2-6,i}=\frac{|m_i|}{2R_{S^2}},
 \qquad
 M_{2-2,ij}=|\sigma_i-\sigma_j|
 =\frac{|m_i-m_j|}{2R_{S^2}}.
 \label{eq:brane-fundamental-adjoint-masses}
\end{equation}
For equal monopole charges, the D2 stack is displaced collectively
relative to the D6-branes. More generally, write the distinct positive
charges as $N_\alpha=\Lambda+\nu_\alpha>0$, with fixed integers $\nu_\alpha$. The
masses are
\begin{equation}
 M_{2-6,\alpha}=\frac{\Lambda+\nu_\alpha}{2R_{S^2}},
 \qquad
 M_{2-2,\alpha\beta}
 =\frac{|\nu_\alpha-\nu_\beta|}{2R_{S^2}}.
 \label{eq:brane-common-shift-masses}
\end{equation}
The common shift increases the D2--D6 separation while preserving
relative D2--D2 separations.

The separation mass is one term in the quadratic fluctuation operator.
To obtain the cylinder frequency, a fundamental scalar of R-charge
$1/2$ must also be expanded in monopole harmonics, whose spins obey
$j=|m_i|/2,|m_i|/2+1,\ldots$. Combining their covariant Laplacian with
the background mass and the conformal curvature coupling gives
\begin{align}
 E_j^2
 &=\frac{j(j+1)-m_i^2/4}{R_{S^2}^2}
   +\frac{m_i^2}{4R_{S^2}^2}
   +\frac{1}{4R_{S^2}^2}
 =\frac{(j+\tfrac12)^2}{R_{S^2}^2},
 \nonumber\\
 E_{\min}^{\mathrm{fund}}
 &=\frac{|m_i|+1}{2R_{S^2}}.
 \label{eq:brane-cylinder-gap}
\end{align}
This is the charged quadratic spectrum entering the localization
calculation \cite{Kim:2009wb,Imamura:2011su}. For an
adjoint-hypermultiplet scalar, $m_i$ is replaced by $m_i-m_j$.

Under the common charge shift $\Lambda \to \infty$, the mass gap of the fundamental
hypermultiplets grows, while the adjoint fluctuation spectrum remains fixed.
Fundamental excitations therefore contribute to the normalized index only
at increasingly high orders, so their contributions vanish at each fixed
order in the large-charge limit. The bounds on finite-charge corrections are derived in
\cref{sec:matching}.

\section{Indices of ADHM and BMN: review}
\label{sec:indices-review}
We review the fixed-vacuum BMN index and the ADHM superconformal
index, with a common choice of charges and fugacities.

\subsection{BMN index}
\label{sec:bmn-review}
\label{sec:bmn-model}
\label{sec:bmn-vacua}
\label{sec:bmn-general-vacuum}
\label{sec:bmn-kernel-general}
\label{sec:bmn-one-block}

BMN matrix quantum mechanics is the maximally supersymmetric mass
deformation of the $U(N_{\BMN})$ BFSS model
\cite{Berenstein:2002jq,Banks:1996vh}. Its nine bosonic matrices split
into an $SO(3)$ triplet $X^i$ and an $SO(6)$ sextet $X^a$. In standard
normalization its Hamiltonian is
\begin{align}
 H={}&R\,\Tr\!\left[
 \frac12\sum_{A=1}^{9}(P^A)^2
 -\frac{1}{4\ell_p^6}\sum_{A,B=1}^{9}[X^A,X^B]^2
 -\frac{1}{2\ell_p^3}\Psi^{\mathsf T}\gamma^A[X^A,\Psi]\right]
 \nonumber\\
 &+\frac{R}{2}\Tr\!\left[
 \left(\frac{\mu}{3R}\right)^2\sum_{i=1}^{3}(X^i)^2
 +\left(\frac{\mu}{6R}\right)^2\sum_{a=4}^{9}(X^a)^2
 +\frac{\ii\mu}{4R}\Psi^{\mathsf T}\gamma^{123}\Psi
 +\frac{2\ii\mu}{3R\ell_p^3}\epsilon_{ijk}X^iX^jX^k
 \right].
 \label{eq:bmn-hamiltonian}
\end{align}
The Myers term stabilizes fuzzy-sphere configurations, while the mass terms
lift the flat directions responsible for the continuum in the undeformed
BFSS model. The resulting quantum mechanics has superalgebra $SU(2|4)$
and discrete oscillator spectra around its isolated classical vacua
\cite{Dasgupta:2002hx,Kim:2002wg,Kim:2002qz}.

Choose the supercharge $\cQ_{\BMN}=Q_{4-}$, with
\begin{equation}
  2\Delta_{\BMN}=\{\cQ_{\BMN},\cQ_{\BMN}^{\dagger}\}
  =2H-\frac{2\mu}{3}M^{12}
   -\frac{\mu}{3}\left(M^{45}+M^{67}+M^{89}\right).
  \label{eq:bmn-bps}
\end{equation}
The index receives contributions from states with $\Delta_{\BMN}=0$.
The three charges
\begin{equation}
 J_a=M^{12}+M^{45},\qquad
 J_b=M^{12}+M^{67},\qquad
 J_c=M^{12}+M^{89}
 \label{eq:bmn-commuting-charges}
\end{equation}
commute with $\cQ_{\BMN}$ and define the refined Witten index
\begin{equation}
  \cI_{\BMN}(a,b,c)
  =\Tr_{\Delta_{\BMN}=0}\!\left[
  (-1)^{2M^{12}}a^{J_a}b^{J_b}c^{J_c}\right].
  \label{eq:bmn-trace}
\end{equation}
We write $Q=abc$. The index is protected under continuous changes of the
coupling for which the discrete gapped description remains valid
\cite{Witten:1982df,Chang:2024zwt}.

The supersymmetric BMN vacua obey
\begin{equation}
  X^a=0,
  \qquad
  X^i=\frac{\mu\ell_p^3}{3R}J^i,
  \qquad
  [J^i,J^j]=\ii\epsilon^{ijk}J^k .
  \label{eq:bmn-vacuum}
\end{equation}
Thus the matrices $J^i$ form a representation of $SU(2)$ and decompose as
in \eqref{eq:bmn-partition-dictionary}. Each irreducible block describes a
fuzzy two-sphere, and $n_\alpha$ identical blocks leave a
$U(n_\alpha)$ gauge factor unbroken. Rectangular fluctuations between blocks
$N_\alpha$ and $N_\beta$ decompose into angular momenta
\begin{equation}
  j=\frac{|N_\alpha-N_\beta|}{2},
  \frac{|N_\alpha-N_\beta|}{2}+1,\ldots,
  \frac{N_\alpha+N_\beta}{2}-1 .
  \label{eq:fuzzy-angular-range}
\end{equation}
The finite angular-momentum range provides the fuzzy-sphere ultraviolet
cutoff. The minimum angular momentum is determined by the difference of
the two block sizes, while the maximum grows with their sum
\cite{Dasgupta:2002ru,Maldacena:2002rb,Lin:2005nh}.

For blocks of sizes $N_\alpha$ and $N_\beta$, define
\begin{equation}
  d_{\alpha\beta}=|N_\alpha-N_\beta|,
  \qquad
  L_{\alpha\beta}=\min(N_\alpha,N_\beta).
  \label{eq:d-and-L}
\end{equation}
The $m$th single-letter index is given by
\begin{equation}
 \iota_{\alpha\beta}^{(m)}
 =\delta_{\alpha\beta}
 +\sum_{j=d_{\alpha\beta}/2}^{(N_\alpha+N_\beta)/2-1}
 (-1)^{2jm+1}Q^{jm}
 (1-a^m)(1-b^m)(1-c^m).
 \label{eq:bmn-letter-spin-sum}
\end{equation}
The block-dependent parity is important at the level of individual
bifundamental letters. It cancels in gauge-invariant words, but retaining
the $m$ dependence gives the correct fermionic plethystic exponential
\cite{Chang:2024zwt}. Summing the geometric progression yields
\begin{equation}
 \iota_{\alpha\beta}^{(m)}
 =\delta_{\alpha\beta}
 -(1-a^m)(1-b^m)(1-c^m)
 (-1)^{md_{\alpha\beta}}Q^{md_{\alpha\beta}/2}
 \frac{1-Q^{mL_{\alpha\beta}}}{1-Q^m}.
 \label{eq:bmn-letter-compact}
\end{equation}

Let $U_\alpha\in U(n_\alpha)$ denote the holonomy of the unbroken group.
The fixed-sector index is
\begin{equation}
 \cI_{\BMN}^{\{n_\alpha,N_\alpha\}}
 =\int\prod_{\alpha=1}^{s}[\dd U_\alpha]\,
 \exp\!\left[
 \sum_{m=1}^{\infty}\frac1m
 \sum_{\alpha,\beta=1}^{s}
 \iota_{\alpha\beta}^{(m)}
 \Tr U_\alpha^{\dagger m}\Tr U_\beta^m
 \right].
 \label{eq:bmn-general-index}
\end{equation}
Here $[\dd U_\alpha]$ is normalized Haar measure. The full BMN index at
matrix size $N_{\BMN}$ is the sum of
\eqref{eq:bmn-general-index} over every partition
$\sum_\alpha n_\alpha N_\alpha=N_{\BMN}$
\cite{Chang:2026allsectors}. We compare the contribution from a fixed BMN
vacuum.

For the contribution at fixed angular momentum, define the kernel
\begin{equation}
 \cK(y)=
 \frac{(1-y)(1-aby)(1-acy)(1-bcy)}
      {(1-ay)(1-by)(1-cy)(1-Qy)} .
 \label{eq:one-level-kernel}
\end{equation}
Since
\begin{equation}
  \log\cK(y)
  =-\sum_{m=1}^{\infty}\frac{y^m}{m}
    (1-a^m)(1-b^m)(1-c^m),
  \label{eq:log-kernel}
\end{equation}
the contribution for each pair of holonomy eigenvalues can be expressed
in terms of $\cK$.
If $u_{\alpha,p}$ are the eigenvalues of $U_\alpha$, then
\begin{equation}
 H_{\alpha\beta}^{\BMN}(x)
 =(1-x)^{-\delta_{\alpha\beta}}
 \prod_{\ell=0}^{L_{\alpha\beta}-1}
 \cK\!\left(
 (-1)^{d_{\alpha\beta}}
 Q^{d_{\alpha\beta}/2+\ell}x\right).
 \label{eq:bmn-general-kernel}
\end{equation}
The allowed angular momenta follow from decomposing the rectangular
fluctuations into irreducible $SU(2)$ representations:
\begin{equation}
 \operatorname{Hom}(\mathbb C^{N_\beta},\mathbb C^{N_\alpha})
 \simeq
 \bigoplus_{\ell=0}^{L_{\alpha\beta}-1}
 \mathcal R_{d_{\alpha\beta}/2+\ell},
 \label{eq:rectangular-decomposition}
\end{equation}
where $\mathcal R_j$ is the spin-$j$ representation.

For $n$ identical blocks of size $N_1$, one has
$d=0$, $L=N_1$ and unbroken group $U(n)$. Writing the holonomies as
$u_i$, the fixed-sector index becomes
\begin{equation}
 \cI_{\BMN}^{n;N_1}
 =\frac1{n!}\oint\prod_{i=1}^{n}\frac{\dd u_i}{2\pi\ii u_i}\,
 \Delta(u)\prod_{i,j=1}^{n}
 H_{N_1}^{\BMN}(u_i/u_j),
 \label{eq:bmn-one-block-index}
\end{equation}
with
\begin{equation}
 \Delta(u)=\prod_{i\neq j}(1-u_i/u_j)
 \label{eq:haar-factor}
\end{equation}
and
\begin{equation}
 H_{N_1}^{\BMN}(x)
 =\frac{1}{1-Q^{N_1}x}
 \frac{
  \PochF{abx}{Q}{N_1}\PochF{acx}{Q}{N_1}
  \PochF{bcx}{Q}{N_1}}
 {\PochF{ax}{Q}{N_1}\PochF{bx}{Q}{N_1}
  \PochF{cx}{Q}{N_1}}.
 \label{eq:bmn-finite-one-block-kernel}
\end{equation}
For $|Q|<1$, the large-block limit, taken order by order in the index
expansion, is
\begin{equation}
 h(x)=\lim_{N_1\to\infty}H_{N_1}^{\BMN}(x)
 =\frac{\Poch{abx}{Q}\Poch{acx}{Q}\Poch{bcx}{Q}}
       {\Poch{ax}{Q}\Poch{bx}{Q}\Poch{cx}{Q}}.
 \label{eq:bmn-infinite-kernel}
\end{equation}
Consequently,
\begin{equation}
 \cI_{\BMN}^{n,\infty}
 =\frac1{n!}\oint\prod_{i=1}^{n}\frac{\dd u_i}{2\pi\ii u_i}\,
 \Delta(u)\prod_{i,j=1}^{n}h(u_i/u_j).
 \label{eq:bmn-large-block-index}
\end{equation}
This is the common matrix integral that will emerge from the equal-GNO
ADHM sector. The finite kernel contains angular levels only up to
$N_1-1$; the limiting kernel contains the complete tower.

\subsection{ADHM index}
\label{sec:adhm-index}
\label{sec:adhm-matter}
\label{sec:adhm-trace}
\label{sec:adhm-cartan-embedding}
\label{sec:adhm-fugacity-dictionary}
\label{sec:adhm-full-formula}
\label{sec:adhm-anatomy}

We use the multiplets and superpotential given in
\cref{tab:adhm-fields} and \eqref{eq:adhm-superpotential}. The manifest global symmetry is
$SU(2)_\ell\times SU(2)_{R_1}\times SU(2)_{R_2}\times
U(1)_T\times U(N_f)$. The first factor rotates the two adjoint chirals,
$SU(2)_{R_1}\times SU(2)_{R_2}$ is the $\mathcal N=4$ R-symmetry, and
$U(1)_T$ is topological. We suppress non-Abelian $U(N_f)$ flavor
fugacities. In terms of the R-symmetry Cartans, the $\mathcal N=2$
superconformal R-charge and the flavor charge used below are
\begin{equation}
 R=r_1+r_2,\qquad F=r_1-r_2.
 \label{eq:R-F-definition}
\end{equation}
The chiral-multiplet charges are collected in
\cref{tab:adhm-charges}, following \cite{Bobev:2022jte}.

\begin{table}[H]
\centering
\renewcommand{\arraystretch}{1.15}
\begin{tabular}{@{}c c c c c c c@{}}
\toprule
$\mathcal N=2$ chiral & $R$ & $\ell$ & $F$ & $r_1$ & $r_2$ & $T$ \\
\midrule
$\Phi_1$ & $\tfrac12$ & $\tfrac12$ & $\tfrac12$ & $\tfrac12$ & $0$ & $0$ \\
$\Phi_2$ & $\tfrac12$ & $-\tfrac12$ & $\tfrac12$ & $\tfrac12$ & $0$ & $0$ \\
$\Phi_3$ & $1$ & $0$ & $-1$ & $0$ & $1$ & $0$ \\
$\psi_q$ & $\tfrac12$ & $0$ & $\tfrac12$ & $\tfrac12$ & $0$ & $0$ \\
$\widetilde\psi_q$ & $\tfrac12$ & $0$ & $\tfrac12$ & $\tfrac12$ & $0$ & $0$ \\
\bottomrule
\end{tabular}
\caption{Charges of the $\mathcal N=2$ chiral multiplets. The topological
symmetry acts on monopole sectors rather than on elementary fields.}
\label{tab:adhm-charges}
\end{table}

Choose the three-dimensional supercharge $\cQ_{\ADHM}$ satisfying
\begin{equation}
 \{\cQ_{\ADHM},\cQ_{\ADHM}^{\dagger}\}
 =\Delta_{3d}-R-j_3.
 \label{eq:adhm-bps-operator}
\end{equation}
Here $\Delta_{3d}$ is the scaling dimension.
States contributing to its index saturate the bound
\begin{equation}
 \Delta_{3d}=R+j_3.
 \label{eq:adhm-bps-condition}
\end{equation}
In the conventions of \cite{Bobev:2022jte}, the refined trace is
\begin{equation}
  {
 \cI_{\ADHM}(q,\xi_\ell,\xi_F,\xi_T)
 =\Tr_{\mathrm{BPS}}\!\left[
 (-1)^{\mathsf F}q^{R+2j_3}
 \xi_\ell^\ell\xi_F^F\xi_T^T\right].}
 \label{eq:ADHM-trace}
\end{equation}
The charges $R+2j_3$, $\ell$, $F$ and $T$ commute with
$\cQ_{\ADHM}$. Here $\mathsf F$ denotes fermion number, while $F$ is
the flavor charge in \eqref{eq:R-F-definition}.

Coulomb-branch localization on $S^2\times S^1$ reduces the path integral
to a sum over integer GNO charge vectors $\bm m\in\bZ^N$ and a contour integral
over unit-circle holonomies $z_i=e^{\ii h_i}$. Each contribution is
computed around the supersymmetric monopole saddle
\eqref{eq:coulomb-monopole-saddle} \cite{Imamura:2011su}.
The vector multiplet contributes the Haar measure deformed by magnetic
root charges. Chiral multiplets contribute ratios of infinite Pochhammer
symbols determined by their R-charges, flavor charges and gauge weights.
The mixed gauge--topological Chern--Simons coupling supplies the factor
$\xi_T^{\sum_i m_i}$.

The ADHM superconformal algebra is $\mathfrak{osp}(4|4)$, with bosonic
subalgebra $\mathfrak{so}(3,2)\oplus\mathfrak{su}(2)_{R_1}
\oplus\mathfrak{su}(2)_{R_2}$. For $N_f=1$, the interacting infrared
theory is expected to exhibit $\mathcal N=8$ enhancement to
$\mathfrak{osp}(8|4)$. The charge comparison below uses only the manifest
Cartans and applies to general fixed $N_f$. On the BMN side, the
bosonic subalgebra of $\mathfrak{su}(2|4)$ is
$\mathfrak{so}(3)\oplus\mathfrak{so}(6)\oplus\mathfrak{u}(1)_H$.
We compare the commuting charges in the two protected traces through
their action on adjoint letters and spherical harmonics.

Define the complex transverse BMN matrices by
$Z_1=X^4+iX^5$, $Z_2=X^6+iX^7$ and $Z_3=X^8+iX^9$.
Their $(M^{45},M^{67},M^{89})$ weights are respectively
$(1,0,0)$, $(0,1,0)$ and $(0,0,1)$. We choose the assignment
\begin{equation}
 \Phi_1\longleftrightarrow Z_1,\qquad
 \Phi_2\longleftrightarrow Z_2,\qquad
 \Phi_3\longleftrightarrow Z_3.
 \label{eq:letter-identification}
\end{equation}
For this assignment, the three independent adjoint weights in
\cref{tab:adhm-charges} uniquely determine the linear map of the
$SO(6)$ Cartans. Matching the magnetic quantum numbers of the $S^2$
harmonics with those of the fuzzy-sphere harmonics identifies the spatial
Cartan $j_3$ with the BMN $SO(3)$ Cartan $M^{12}$. Together these give
\begin{equation}
 M^{12}=j_3,\qquad
 M^{45}=r_1+\ell,\qquad
 M^{67}=r_1-\ell,\qquad
 M^{89}=r_2.
 \label{eq:cartan-embedding}
\end{equation}
Permuting \eqref{eq:letter-identification} permutes the BMN fugacities
$(a,b,c)$. This identification fixes the charge convention for the index
comparison.
Substituting \eqref{eq:cartan-embedding} into
\eqref{eq:bmn-commuting-charges} gives
\begin{equation}
 J_a=j_3+r_1+\ell,\qquad
 J_b=j_3+r_1-\ell,\qquad
 J_c=j_3+r_2.
 \label{eq:J-in-adhm-charges}
\end{equation}
The corresponding fugacity weights in the two traces are
\begin{align}
 w_{\BMN}
 &=(abc)^{j_3}(ab)^{r_1}c^{r_2}(a/b)^\ell,
 \label{eq:bmn-weight-collected}\\
 w_{\ADHM}
 &=(q^2)^{j_3}(q\xi_F)^{r_1}
   (q\xi_F^{-1})^{r_2}\xi_\ell^\ell.
 \label{eq:adhm-weight-collected}
\end{align}
Matching these weights fixes
\begin{equation}
 a=q^{1/2}\xi_\ell^{1/2}\xi_F^{1/2},\qquad
 b=q^{1/2}\xi_\ell^{-1/2}\xi_F^{1/2},\qquad
 c=q\xi_F^{-1},
 \label{eq:adhm-to-bmn-map}
\end{equation}
so
\begin{equation}
 Q=abc=q^2.
 \label{eq:Q-q-relation}
\end{equation}
The inverse map is
\begin{equation}
 q=Q^{1/2},\qquad
 \xi_\ell=\frac{a}{b},\qquad
 \xi_F=\left(\frac{ab}{c}\right)^{1/2}.
 \label{eq:bmn-to-adhm-map}
\end{equation}
The half-integer powers follow the original ADHM fugacity convention
and are treated as formal powers in the index expansion.

Substituting the charges in \cref{tab:adhm-charges} into the general
three-dimensional localization formula gives
\begingroup
\footnotesize
\begin{align}
\cI_{\ADHM}(q,\xi_\ell,\xi_F,\xi_T)
={}&\frac1{N!}\sum_{\bm m\in\bZ^N}\oint
 \left(\prod_{i=1}^{N}\frac{\dd z_i}{2\pi\ii z_i}\,\xi_T^{m_i}\right)
 \prod_{i\neq j}q^{-\frac12|m_i-m_j|}
 \left(1-z_i z_j^{-1}q^{|m_i-m_j|}\right)
 \nonumber\\
&\times\prod_{i,j=1}^{N}
 \left(q^{1/2}z_i^{-1}z_j\xi_\ell^{-1/2}\xi_F^{-1/2}\right)^{\frac12|m_i-m_j|}
 \frac{
  \Poch{z_i^{-1}z_j\xi_\ell^{-1/2}\xi_F^{-1/2}
    q^{3/2+|-m_i+m_j|}}{q^2}}
 {\Poch{z_i z_j^{-1}\xi_\ell^{1/2}\xi_F^{1/2}
    q^{1/2+|m_i-m_j|}}{q^2}}
 \nonumber\\
&\times\prod_{i,j=1}^{N}
 \left(q^{1/2}z_i^{-1}z_j\xi_\ell^{1/2}\xi_F^{-1/2}\right)^{\frac12|m_i-m_j|}
 \frac{
  \Poch{z_i^{-1}z_j\xi_\ell^{1/2}\xi_F^{-1/2}
    q^{3/2+|-m_i+m_j|}}{q^2}}
 {\Poch{z_i z_j^{-1}\xi_\ell^{-1/2}\xi_F^{1/2}
    q^{1/2+|m_i-m_j|}}{q^2}}
 \nonumber\\
&\times\prod_{i,j=1}^{N}
 \left(z_i^{-1}z_j\xi_F\right)^{\frac12|m_i-m_j|}
 \frac{\Poch{z_i^{-1}z_j\xi_Fq^{1+|-m_i+m_j|}}{q^2}}
      {\Poch{z_i z_j^{-1}\xi_F^{-1}q^{1+|m_i-m_j|}}{q^2}}
 \nonumber\\
&\times\prod_{i=1}^{N}\left[
 \left(q^{1/2}z_i^{-1}\xi_F^{-1/2}\right)^{\frac12|m_i|}
 \frac{\Poch{z_i^{-1}\xi_F^{-1/2}q^{3/2+|-m_i|}}{q^2}}
      {\Poch{z_i\xi_F^{1/2}q^{1/2+|m_i|}}{q^2}}
 \right.
 \nonumber\\[-2pt]
&\hspace{22mm}\left.
 \times
 \left(q^{1/2}z_i\xi_F^{-1/2}\right)^{\frac12|-m_i|}
 \frac{\Poch{z_i\xi_F^{-1/2}q^{3/2+|m_i|}}{q^2}}
      {\Poch{z_i^{-1}\xi_F^{1/2}q^{1/2+|-m_i|}}{q^2}}
 \right]^{N_f}.
 \label{eq:ADHM-localized}
\end{align}
\endgroup

We use
\begin{equation}
  \Poch{x}{Q}=\prod_{r=0}^{\infty}(1-xQ^r),
  \qquad |Q|<1,
  \label{eq:pochhammer-definition}
\end{equation}
with $Q=q^2$ as in \eqref{eq:Q-q-relation}. The localized integrand factors as
\begin{equation}
 \begin{aligned}
 \cI_{\ADHM}
 ={}&\frac1{N!}\sum_{\bm m\in\bZ^N}
 \oint\prod_{i=1}^{N}
 \left(\frac{\dd z_i}{2\pi iz_i}\,\xi_T^{m_i}\right)
 \cZ_{\mathrm{vec}}(z,\bm m)
 \prod_{A=1}^{3}\cZ_{\Phi_A}(z,\bm m)
 \prod_{i=1}^{N}\cZ_{\mathrm{hyp}}(z_i,m_i)^{N_f}.
 \end{aligned}
\label{eq:adhm-factorized}
\end{equation}
The vector and adjoint factors depend on
$d_{ij}=|m_i-m_j|$, whereas a fundamental factor depends on $|m_i|$.
This distinction is the algebraic origin of the equal-GNO simplification.

After grouping charge vectors related by permutations, each monopole-sector
contribution carries $1/\prod_\alpha n_\alpha!$, where $n_\alpha$ is the
multiplicity of each distinct charge. This factor accounts for permutations
among eigenvalues with equal charge. For equal charges it reduces to
$1/N!$. The factor $\xi_T^{\sum_i m_i}$ records the total topological
charge and is common to the whole sector.

\section{ADHM--BMN index matching}
\label{sec:matching}

\subsection{Equal charges}
\label{sec:equal-matching}
\label{sec:fugacity-map}
\label{sec:step1}
\label{sec:step2}
\label{sec:step3}
\label{sec:step4}
\label{sec:step5}
\label{sec:magnetic-gap}
In the equal-charge monopole sector, the vector and adjoint determinants reduce to
the infinite BMN kernel. The fundamental determinant supplies the factor
$c^{NN_fN_1/2}$, with $c=q\xi_F^{-1}$. The product
$\xi_T^{NN_1}c^{NN_fN_1/2}$ is the contribution of the bare monopole.
The remaining oscillator factor tends to one at each fixed order in $q$
as the common charge grows.

Using \eqref{eq:adhm-to-bmn-map} with $Q=q^2$, we have
\begin{equation}
  abc=Q,\qquad
  ab=q\xi_F,\qquad
  ac=q^{3/2}\xi_\ell^{1/2}\xi_F^{-1/2},\qquad
  bc=q^{3/2}\xi_\ell^{-1/2}\xi_F^{-1/2}.
  \label{eq:map-products}
\end{equation}
The fixed longitudinal momentum contributes a common power of $\xi_T$,
which is removed when normalizing the sector.

In \eqref{eq:adhm-factorized}, choose
\begin{equation}
  m_i=N_1,\qquad i=1,\ldots,N .
  \label{eq:positive-equal-sector}
\end{equation}
Then
\begin{equation}
  d_{ij}=0,\qquad |m_i|=N_1,\qquad
  \prod_{i=1}^{N}\xi_T^{m_i}=\xi_T^{NN_1}.
  \label{eq:positive-sector-data}
\end{equation}
All $N!$ permutations leave this equal-charge vector unchanged, so the
contour integral retains the factor $1/N!$.

Setting $d_{ij}=0$ in the vector contribution gives
\begin{equation}
 Z_{\mathrm{vec}}(z,(N_1)^N)
 =\prod_{i\neq j}(1-z_i z_j^{-1})
 \equiv\Delta(z).
 \label{eq:vector-haar}
\end{equation}
This is the $U(N)$ Haar factor.
Put $u_{ij}=z_i/z_j$. At $d_{ij}=0$, the adjoint zero-point factors
in \eqref{eq:ADHM-localized} are
\[
 \begin{aligned}
 \left(q^{1/2}u_{ij}^{-1}\xi_\ell^{\mp1/2}\xi_F^{-1/2}\right)^{d_{ij}/2}
 &=1 &&\text{for }\Phi_{1,2},\\
 \left(u_{ij}^{-1}\xi_F\right)^{d_{ij}/2}
 &=1 &&\text{for }\Phi_3.
 \end{aligned}
\]
Using \eqref{eq:adhm-to-bmn-map} and \eqref{eq:map-products}, the
denominators of the three adjoint determinants are
\begin{equation}
  \Poch{au_{ij}}{Q},\qquad
  \Poch{bu_{ij}}{Q},\qquad
  \Poch{cu_{ij}}{Q},
  \label{eq:adjoint-denominators}
\end{equation}
and their numerators are
\begin{equation}
  \Poch{bc/u_{ij}}{Q},\qquad
  \Poch{ac/u_{ij}}{Q},\qquad
  \Poch{ab/u_{ij}}{Q}.
  \label{eq:adjoint-numerators}
\end{equation}
Relabelling $i\leftrightarrow j$ in the product gives
\begin{equation}
 \prod_{i,j=1}^{N}\Poch{\alpha/u_{ij}}{Q}
 =\prod_{i,j=1}^{N}\Poch{\alpha u_{ij}}{Q}.
 \label{eq:pair-inversion}
\end{equation}
It follows that
\begin{equation}
 \prod_{A=1}^{3}Z_{\Phi_A}(z,(N_1)^N)
 =\prod_{i,j=1}^{N}h(z_i/z_j).
 \label{eq:adjoint-exact-match}
\end{equation}
The vector and adjoint reductions hold at every $N_1$.

For one fundamental hypermultiplet at one eigenvalue, substitute
$|m|=N_1$ into the last two lines of \eqref{eq:ADHM-localized}. Separate
the fundamental and antifundamental pieces as
\begin{align}
 F_{N_1}^{(+)}(z)
 ={}&
 \left(q^{1/2}z^{-1}\xi_F^{-1/2}\right)^{N_1/2}
 \frac{\Poch{z^{-1}\xi_F^{-1/2}q^{N_1+3/2}}{Q}}
      {\Poch{z\xi_F^{1/2}q^{N_1+1/2}}{Q}},
 \label{eq:F-plus}\\
 F_{N_1}^{(-)}(z)
 ={}&
 \left(q^{1/2}z\xi_F^{-1/2}\right)^{N_1/2}
 \frac{\Poch{z\xi_F^{-1/2}q^{N_1+3/2}}{Q}}
      {\Poch{z^{-1}\xi_F^{1/2}q^{N_1+1/2}}{Q}}.
 \label{eq:F-minus}
\end{align}
The zero-point factors preceding the Pochhammer ratios multiply to
\begin{align}
 &\left(q^{1/2}z^{-1}\xi_F^{-1/2}\right)^{N_1/2}
  \left(q^{1/2}z\xi_F^{-1/2}\right)^{N_1/2}
 \nonumber\\
 &\hspace{30mm}
 =q^{N_1/2}\xi_F^{-N_1/2}
 =c^{N_1/2}.
 \label{eq:fund-zero-point}
\end{align}
The holonomy powers cancel exactly. Hence one complete hypermultiplet has
the finite-$N_1$ factorization
\begin{equation}
 F_{N_1}(z)
 \equiv F_{N_1}^{(+)}(z)F_{N_1}^{(-)}(z)
 =c^{N_1/2}R_{N_1}(z),
 \label{eq:fund-factorization}
\end{equation}
where
\begin{equation}
 R_{N_1}(z)=
 \frac{
 \Poch{z^{-1}\xi_F^{-1/2}q^{N_1+3/2}}{Q}
 \Poch{z\xi_F^{-1/2}q^{N_1+3/2}}{Q}}
 {\Poch{z\xi_F^{1/2}q^{N_1+1/2}}{Q}
  \Poch{z^{-1}\xi_F^{1/2}q^{N_1+1/2}}{Q}}.
 \label{eq:R-exact}
\end{equation}
For a small argument $\varepsilon$, the definition
\eqref{eq:pochhammer-definition} gives
\begin{equation}
  \Poch{\varepsilon}{Q}
  =1-\frac{\varepsilon}{1-Q}+\cO(\varepsilon^2),
  \qquad
  \frac{1}{\Poch{\varepsilon}{Q}}
  =1+\frac{\varepsilon}{1-Q}+\cO(\varepsilon^2).
  \label{eq:small-pochhammer}
\end{equation}
Defining
\begin{equation}
\rho_{N_1}=q^{N_1+\frac{1}{2}},
\end{equation}
the four small arguments of \eqref{eq:R-exact} are
\begin{equation}
q\rho_{N_1}z^{-1}\xi_F^{-1/2},
\qquad
q\rho_{N_1}z\xi_F^{-1/2},
\qquad
\rho_{N_1}z\xi_F^{1/2},
\qquad
\rho_{N_1}z^{-1}\xi_F^{1/2}.
\label{eq:four-arguments}
\end{equation}
Expanding the fundamental oscillator factor in \eqref{eq:R-exact} gives
\begin{equation}
 R_{N_1}(z)=1+
 \frac{q^{N_1+1/2}(\xi_F^{1/2}-q\xi_F^{-1/2})}{1-q^2}
 (z+z^{-1})+\cO(q^{2N_1+1}).
 \label{eq:R-first-correction}
\end{equation}
For fixed $N$ and $N_f$, the normalized product therefore satisfies
\begin{align}
 c^{-NN_fN_1/2}\prod_{i=1}^{N}F_{N_1}(z_i)^{N_f}
 ={}&1+\frac{N_fq^{N_1+1/2}(\xi_F^{1/2}-q\xi_F^{-1/2})}{1-q^2}
 \sum_{i=1}^{N}(z_i+z_i^{-1})
 \nonumber\\
 &+\cO(q^{2N_1+1}).
 \label{eq:all-fundamentals}
\end{align}
At any fixed order $q^d$, the correction terms cannot contribute once
$N_1+1/2>d$. Hence
\begin{equation}
 \lim_{N_1\to\infty}
 c^{-NN_fN_1/2}\prod_{i=1}^{N}F_{N_1}(z_i)^{N_f}=1
 \label{eq:fundamental-limit}
\end{equation}
order by order in $q$, with the flavor fugacities held as formal variables.

The monopole charge raises the lowest fundamental monopole harmonic to
$j=N_1/2$, so the first fundamental letter contributes at order $q^{N_1+1/2}$.
The adjoint weights depend on $m_i-m_j=0$ and are independent of the
common flux. Thus the oscillator factor in
\eqref{eq:fundamental-limit} tends to one, while the monopole weight
$c^{NN_fN_1/2}$ remains in the unnormalized index. Gauge projection
removes the term linear in $z_i^{\pm1}$ in \eqref{eq:all-fundamentals};
the resulting fundamental correction to the normalized index starts no
earlier than $q^{2N_1+1}$, as shown more generally in
\eqref{eq:general-integrated-bounds}.

Combining \eqref{eq:positive-sector-data}, \eqref{eq:vector-haar},
\eqref{eq:adjoint-exact-match} and
\eqref{eq:fund-factorization}, the exact contribution of the positive
equal-charge monopole sector is
\begin{equation}
 \cI^{\ADHM}_{N,(N_1)^N}
 =\xi_T^{NN_1}\frac1{N!}
 \oint\prod_{i=1}^{N}\frac{\dd z_i}{2\pi\ii z_i}\,
 \Delta(z)\prod_{i,j=1}^{N}h(z_i/z_j)
 \prod_{i=1}^{N}F_{N_1}(z_i)^{N_f}.
 \label{eq:exact-positive-sector}
\end{equation}
Define the normalized fluctuation index
\begin{equation}
 \widehat\cI^{\ADHM}_{N,N_1;N_f}
 \equiv
 \xi_T^{-NN_1}c^{-NN_fN_1/2}
 \cI^{\ADHM}_{N,(N_1)^N}.
 \label{eq:normalized-positive-sector}
\end{equation}
The fundamental limit \eqref{eq:fundamental-limit} gives
\begin{equation}
 \lim_{N_1\to\infty}\widehat\cI^{\ADHM}_{N,N_1;N_f}
 =\frac1{N!}\oint\prod_{i=1}^{N}\frac{\dd z_i}{2\pi\ii z_i}\,
 \Delta(z)\prod_{i,j=1}^{N}h(z_i/z_j).
 \label{eq:adhm-common-integral}
\end{equation}
Identifying the ADHM rank with the BMN block multiplicity, $N=n$, the
right-hand side is \eqref{eq:bmn-large-block-index}. Thus, at fixed $n$
and $N_f$, with the fugacities related by \eqref{eq:adhm-to-bmn-map},
we obtain
\begin{equation}
  {
 \lim_{N_1\to\infty}
 \xi_T^{-nN_1}c^{-nN_fN_1/2}
 \cI^{\ADHM}_{n,(N_1)^n}
 =
 \lim_{N_1\to\infty}\cI_{\BMN}^{n;N_1}
 =
 \cI_{\BMN}^{n,\infty}.}
 \label{eq:main-matching}
\end{equation}
The equality holds order by order in $q$, retaining the flavor fugacities.

The factor $\xi_T^{nN_1}$ records the positive M-circle momentum, while
$c^{nN_fN_1/2}=(q\xi_F^{-1})^{nN_fN_1/2}$ accounts for the bare
monopole's scaling dimension and flavor charge.

At finite $N_1$, the ADHM sector contains the fundamental oscillator
factor $R_{N_1}$, and the BMN kernel has a finite angular cutoff.
Equation~\eqref{eq:main-matching} follows when both corrections move
beyond every fixed order in the index expansion.

\subsection{General charges}
\label{sec:general-placeholder}
\label{sec:general-normalization}
\label{sec:general-monopole-kernel}
\label{sec:general-bmn-tail}

The equal-charge result extends to several distinct positive monopole charges. In this case the vector--adjoint determinant retains the charge
differences, and the harmonics of fields connecting different blocks start
at nonzero spin. The same minimum angular momentum occurs for rectangular
fluctuations around a general BMN vacuum. We derive the exact
finite-charge comparison first, and then take a common large shift of all
block sizes.

Fix the monopole sector represented by \eqref{eq:general-positive-orbit},
with distinct
$N_\alpha>0$ and fixed multiplicities $n_\alpha$. We retain the notation
$d_{\alpha\beta}=|N_\alpha-N_\beta|$ and
$L_{\alpha\beta}=\min(N_\alpha,N_\beta)$ from
\eqref{eq:d-and-L}.
The rank and total charge are
\begin{equation}
 N=\sum_\alpha n_\alpha,
 \qquad
 P=\sum_\alpha n_\alpha N_\alpha=N_{\BMN}.
 \label{eq:general-rank-momentum}
\end{equation}
There are $N!/\prod_\alpha n_\alpha!$ distinct permutations of this
charge vector. Their contributions to \eqref{eq:ADHM-localized} agree
after relabelling the holonomies. Summing them with the prefactor $1/N!$
gives the factor $1/\prod_\alpha n_\alpha!$ accounting for permutations
within each group of equal charges. Relabelling the holonomies as
$z_{\alpha,i}$, define the normalized measure of
$H=\prod_\alpha U(n_\alpha)$ by
\begin{equation}
 [\dd z]_H=
 \frac{1}{\prod_\alpha n_\alpha!}
 \prod_{\alpha,i}\frac{\dd z_{\alpha,i}}{2\pi\ii z_{\alpha,i}}
 \prod_\alpha\prod_{i\ne j}
 \left(1-\frac{z_{\alpha,i}}{z_{\alpha,j}}\right).
 \label{eq:general-haar-measure}
\end{equation}
There is no Haar factor connecting different flux blocks: those roots
carry nonzero monopole charge and remain in the fluctuation kernel.

For two distinct eigenvalues with $d=|m_i-m_j|$, the vector zero-point
factors multiply to $q^{-d}$. The corresponding adjoint zero-point
factors for the ordered pairs $(i,j)$ and $(j,i)$ are
\begin{equation}
 Z^{(ij,ji)}_{\Phi_1,0}=q^{d/2}\xi_\ell^{-d/2}\xi_F^{-d/2},
 \quad
 Z^{(ij,ji)}_{\Phi_2,0}=q^{d/2}\xi_\ell^{d/2}\xi_F^{-d/2},
 \quad
 Z^{(ij,ji)}_{\Phi_3,0}=\xi_F^{d}.
 \label{eq:general-zero-point-cancellation}
\end{equation}
The opposite holonomy powers have already cancelled between the two
orientations. The vector and adjoint vacuum contributions cancel:
$q^{-d}Z^{(ij,ji)}_{\Phi_1,0}Z^{(ij,ji)}_{\Phi_2,0}
Z^{(ij,ji)}_{\Phi_3,0}=1$.
Each fundamental hypermultiplet instead contributes $c^{N_\alpha/2}$
per eigenvalue, exactly as in \eqref{eq:fund-zero-point}. The complete
bare-monopole contribution and normalized sector are consequently
\begin{equation}
 W_{\bm m}=\xi_T^{P}c^{N_fP/2},
 \qquad
 \widehat\cI_{\ADHM}^{\bm m;N_f}
 =W_{\bm m}^{-1}\cI_{\ADHM}^{\bm m;N_f}.
 \label{eq:general-sector-normalization}
\end{equation}
For $N_f=1$, $W_{\bm m}=\lambda^P$ with
$\lambda=c^{1/2}\xi_T$.

For the vector--adjoint contribution, set
$x=z_{\beta,j}/z_{\alpha,i}$ for an ordered pair. Inverting the pair
labels in the chiral numerators, as in \eqref{eq:pair-inversion}, leaves
the charge difference unchanged. Using \eqref{eq:adhm-to-bmn-map}, the
vector--adjoint determinant becomes the residual Haar factor times the
ordered-pair kernel
\begin{equation}
 \widetilde H^{\mathrm{mono}}_{\alpha\beta}(x)
 =\frac{1-Q^{d_{\alpha\beta}/2}x}{(1-x)^{\delta_{\alpha\beta}}}
 \frac{
  \Poch{abQ^{d_{\alpha\beta}/2}x}{Q}
  \Poch{acQ^{d_{\alpha\beta}/2}x}{Q}
  \Poch{bcQ^{d_{\alpha\beta}/2}x}{Q}}
 {\Poch{aQ^{d_{\alpha\beta}/2}x}{Q}
  \Poch{bQ^{d_{\alpha\beta}/2}x}{Q}
  \Poch{cQ^{d_{\alpha\beta}/2}x}{Q}}.
 \label{eq:general-monopole-pochhammer}
\end{equation}
For $\alpha=\beta$, the factors $(1-x)$ cancel before evaluating the
kernel at $x=1$. Thus
$\widetilde H^{\mathrm{mono}}_{\alpha\alpha}(x)=h(x)$, including the
Cartan contribution $h(1)$. For different eigenvalues in the same block,
the vector root is supplied by \eqref{eq:general-haar-measure}; for
$\alpha\ne\beta$, it is the numerator
$1-Q^{d_{\alpha\beta}/2}x$ in
\eqref{eq:general-monopole-pochhammer}.

Using the contribution at fixed angular momentum in
\eqref{eq:one-level-kernel}, we can equivalently write a product over the
allowed angular momenta:
\begin{equation}
 \widetilde H^{\mathrm{mono}}_{\alpha\beta}(x)
 =(1-x)^{-\delta_{\alpha\beta}}
 \prod_{\ell=0}^{\infty}
 \cK\!\left(Q^{d_{\alpha\beta}/2+\ell}x\right).
 \label{eq:general-monopole-harmonics}
\end{equation}
Indeed, the factors $(1-y)/(1-Qy)$ telescope to
$1-Q^{d_{\alpha\beta}/2}x$, and the remaining factors give the six
Pochhammer symbols in \eqref{eq:general-monopole-pochhammer}. The lower
spin $d_{\alpha\beta}/2$ is the monopole-harmonic bound for a field of
monopole charge $N_\alpha-N_\beta$; unlike a finite rectangular matrix,
the continuum sphere has no upper angular cutoff.

For any factor $\mathcal F(z)$ in the integrand, write
\begin{equation}
 \big\langle\mathcal F\big\rangle_{\mathrm{mono}}
 =\oint[\dd z]_H
 \prod_{\alpha,\beta}\prod_{i=1}^{n_\alpha}\prod_{j=1}^{n_\beta}
 \widetilde H^{\mathrm{mono}}_{\alpha\beta}
 \!\left(\frac{z_{\beta,j}}{z_{\alpha,i}}\right)
 \mathcal F(z).
 \label{eq:general-common-functional}
\end{equation}
This notation denotes an unnormalized integral, so
$\langle1\rangle_{\mathrm{mono}}$ is the common fluctuation index.
The normalized ADHM sector is exactly
\begin{equation}
 \widehat\cI_{\ADHM}^{\bm m;N_f}
 =\left\langle
   \prod_{\alpha,i}R_{N_\alpha}(z_{\alpha,i})^{N_f}
  \right\rangle_{\mathrm{mono}},
 \label{eq:general-adhm-exact}
\end{equation}
where $R_{N_\alpha}$ is \eqref{eq:R-exact} with $N_1$ replaced by
$N_\alpha$.

The BMN kernel \eqref{eq:bmn-general-kernel} includes
$(-1)^{d_{\alpha\beta}}$. To compare it with the natural ADHM convention,
perform the blockwise change of variables
\begin{equation}
 u_{\alpha,i}=(-1)^{N_\alpha}z_{\alpha,i},
 \qquad
 (-1)^{d_{\alpha\beta}}
 \frac{u_{\beta,j}}{u_{\alpha,i}}
 =\frac{z_{\beta,j}}{z_{\alpha,i}}.
 \label{eq:general-parity-rephasing}
\end{equation}
The second equality follows because
$|N_\alpha-N_\beta|$ and $N_\beta-N_\alpha$ have the same parity. The
unit circles, differentials and within-block Haar factors are invariant.
With this parity sign absorbed into the holonomies, the BMN integral uses
the same $z$ variables and the finite kernel
\begin{equation}
 \widetilde H^{\BMN}_{\alpha\beta}(x)
 =(1-x)^{-\delta_{\alpha\beta}}
 \prod_{\ell=0}^{L_{\alpha\beta}-1}
 \cK\!\left(Q^{d_{\alpha\beta}/2+\ell}x\right).
 \label{eq:general-bmn-unsigned}
\end{equation}
We have changed variables only in the BMN integral, so the fundamental
factors in \eqref{eq:general-adhm-exact} retain their original arguments.
If both integrals are instead expressed in $u$, those factors must be
written as $R_{N_\alpha}((-1)^{N_\alpha}u_{\alpha,i})$.

The harmonics above the BMN cutoff contribute the factor
\begin{equation}
 T_{\alpha\beta}(x)
 =\prod_{\ell=L_{\alpha\beta}}^{\infty}
   \cK\!\left(Q^{d_{\alpha\beta}/2+\ell}x\right),
 \qquad
 \widetilde H^{\mathrm{mono}}_{\alpha\beta}(x)
 =\widetilde H^{\BMN}_{\alpha\beta}(x)T_{\alpha\beta}(x).
 \label{eq:general-exact-tail-ratio}
\end{equation}
Together with \eqref{eq:general-adhm-exact}, this gives an exact
finite-charge comparison in a single common measure:
\begin{align}
 \widehat\cI_{\ADHM}^{\bm m;N_f}
 &=\left\langle\prod_{\alpha,i}
       R_{N_\alpha}(z_{\alpha,i})^{N_f}\right\rangle_{\mathrm{mono}},
 \nonumber\\
 \cI_{\BMN}^{\{n_\alpha,N_\alpha\}}
 &=\left\langle\prod_{\alpha,\beta,i,j}
       T_{\alpha\beta}(z_{\beta,j}/z_{\alpha,i})^{-1}
    \right\rangle_{\mathrm{mono}}.
 \label{eq:general-finite-comparison}
\end{align}
The leading order and sign of the high-spin contribution
$T_{\alpha\beta}$ follow directly from
\eqref{eq:log-kernel}:
\begin{equation}
 \log T_{\alpha\beta}(x)
 =-\sum_{m=1}^{\infty}\frac{x^m}{m}
  (1-a^m)(1-b^m)(1-c^m)
  \frac{Q^{m(d_{\alpha\beta}/2+L_{\alpha\beta})}}{1-Q^m}.
 \label{eq:general-tail-logarithm}
\end{equation}
Since $d_{\alpha\beta}/2+L_{\alpha\beta}
=(N_\alpha+N_\beta)/2$, the correction can first appear at order
$q^{N_\alpha+N_\beta}$. It comes from the high-spin harmonics above the
range in \eqref{eq:rectangular-decomposition}; the low-spin spectrum is
shared by both theories.

\subsection{Large-charge limit}
\label{sec:general-common-shift}

Choose fixed integers $\nu_\alpha$ and take
\begin{equation}
 N_\alpha(\Lambda)=\Lambda+\nu_\alpha>0,
 \qquad \Lambda\longrightarrow\infty.
 \label{eq:general-common-shift}
\end{equation}
Blocks with the same monopole charge are grouped together. The unbroken gauge group
and all differences $d_{\alpha\beta}=|\nu_\alpha-\nu_\beta|$ then remain
fixed. The integration measure and vector--adjoint determinant in
\eqref{eq:general-common-functional} are therefore independent of
$\Lambda$, while the monopole charges seen by fundamental fields and the
BMN angular cutoffs grow.

We expand formally in $q^{1/2}$, retaining both flavor fugacities. At each
order, the coefficient is a finite sum of flavor weights with integer
powers of $\xi_\ell^{1/2}$ and $\xi_F^{1/2}$, which may be positive or
negative. The holonomy integrals extract the constant term in every
holonomy after including the Haar factor. Writing
$N_{\min}=\min_\alpha N_\alpha$, the explicit products give
\begin{equation}
 \prod_{\alpha,i}R_{N_\alpha}(z_{\alpha,i})^{N_f}
 =1+\cO(q^{N_{\min}+1/2}),
 \qquad
 \prod_{\alpha,\beta,i,j}T_{\alpha\beta}(z_{\beta,j}/z_{\alpha,i})^{-1}
 =1+\cO(q^{2N_{\min}}).
 \label{eq:general-integrand-bounds}
\end{equation}
These bounds hold for fixed multiplicities and fixed $N_f$. The
continuum kernels contain only nonnegative powers of $q$, so multiplying
by them and projecting onto gauge singlets cannot lower either bound. In fact, the first bound
improves after integration: the continuum measure is invariant under
$z_{\alpha,i}\mapsto e^{\ii\theta}z_{\alpha,i}$ for all eigenvalues,
whereas every fundamental letter has center charge $+1$ or $-1$.
At least two such letters are needed. Consequently
\begin{align}
 \widehat\cI_{\ADHM}^{\bm m;N_f}
  -\langle1\rangle_{\mathrm{mono}}
 &=\cO(q^{2N_{\min}+1}),
 \nonumber\\
 \cI_{\BMN}^{\{n_\alpha,N_\alpha\}}
  -\langle1\rangle_{\mathrm{mono}}
 &=\cO(q^{2N_{\min}}).
 \label{eq:general-integrated-bounds}
\end{align}
Corrections cannot appear before these orders; further cancellations may
delay their onset.

The two expressions in \eqref{eq:general-finite-comparison} have the same
continuum measure, which is independent of $\Lambda$. At any fixed order
in $q$, both the fundamental determinant and the product of
$T_{\alpha\beta}^{-1}$ reduce to one when the common monopole charge is
sufficiently large. Since only finitely many holonomy weights contribute
at that order,
the gauge projection preserves this limit. Thus, for fixed
$n_\alpha$, fixed integers $\nu_\alpha$, and fixed $N_f\geq1$, we obtain
\begin{equation}
  {
 \lim_{\Lambda\to\infty}
  \xi_T^{-P_\Lambda}c^{-N_fP_\Lambda/2}
  \cI_{\ADHM}^{\bm m_\Lambda;N_f}
 =\lim_{\Lambda\to\infty}
  \cI_{\BMN}^{\{n_\alpha,\Lambda+\nu_\alpha\}}
 =\langle1\rangle_{\mathrm{mono}}.}
 \label{eq:general-matching}
\end{equation}
Here $\bm m_\Lambda$ is the charge vector in which each positive charge
$N_\alpha(\Lambda)$ from \eqref{eq:general-common-shift} occurs
$n_\alpha$ times, and
$P_\Lambda=\sum_\alpha n_\alpha(\Lambda+\nu_\alpha)$.
The fugacity map is \eqref{eq:adhm-to-bmn-map}. Both limits are understood
order by order in $q$. More precisely,
\eqref{eq:general-integrated-bounds} shows that the two normalized indices
already agree at all orders below $q^{2N_{\min}}$ at finite $\Lambda$.
The mass gap of the fundamental hypermultiplets pushes their excitations
to higher orders, while the growing BMN cutoff postpones the effect of the missing high-spin
harmonics. The relative monopole charges and the low-spin vector--adjoint
spectrum remain fixed.

\section{Conclusions and future directions}
\label{sec:conclusions}

We have established the large-charge index matching between ADHM
monopole sectors and BMN fuzzy-sphere vacua in \eqref{eq:general-matching}.
After dividing the ADHM contribution by the bare-monopole factor
$\xi_T^{P_\Lambda}c^{N_fP_\Lambda/2}$, the indices agree order by order
in $q$ as all positive monopole charges grow with fixed differences,
multiplicities and $N_f$. Fundamental excitations move to higher orders
and the BMN angular-momentum cutoffs grow without bound, leaving the
same vector--adjoint contribution. The bounds on finite-charge
corrections are given in \eqref{eq:general-integrated-bounds}.

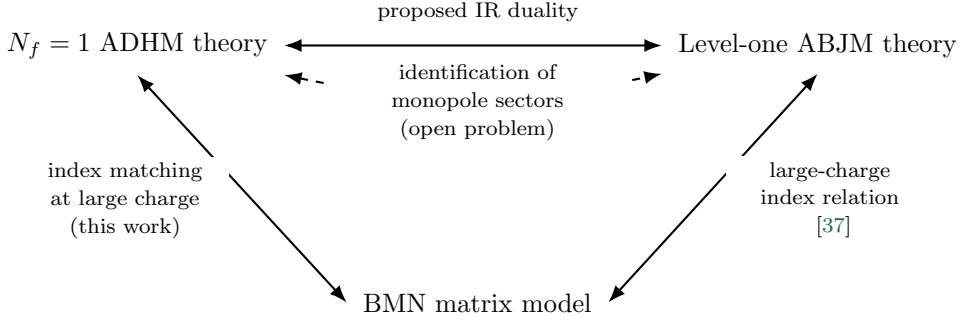
\begin{figure}[htbp]
 \centering
 \begin{tikzpicture}[
   theory/.style={align=center,inner sep=6pt,font=\small},
   relation/.style={<->,>=Latex,thick},
   relabel/.style={fill=white,align=center,font=\scriptsize,inner sep=2pt}]
  \node[theory] (adhm) at (-4.5,1.4) {$N_f=1$ ADHM theory};
  \node[theory] (other) at (4.5,1.4) {Level-one ABJM theory};
  \node[theory] (bmn) at (0,-2) {BMN matrix model};
  \draw[relation] (adhm.east) -- (other.west);
  \node[relabel,text width=3.6cm] at (0,1.85) {proposed IR duality};
  \draw[relation,dashed] (adhm.south east) to[bend right=12] (other.south west);
  \node[relabel,text width=3.8cm] at (0,0.65) {identification of monopole sectors\\(open problem)};
  \draw[relation] (adhm.south) --
   node[relabel,pos=0.55,left=3mm,text width=2.6cm]
   {index matching at large charge\\(this work)} (bmn.west);
  \draw[relation] (other.south) --
   node[relabel,pos=0.55,right=3mm,text width=2.6cm]
   {large-charge index relation\\\cite{Chang:2026yeg}} (bmn.east);
 \end{tikzpicture}
 \caption{For $N_f=1$, the proposed ADHM--ABJM infrared duality and the two large-charge index
 relations. The dashed edge denotes the unresolved map between the
 selected sectors.}
 \label{fig:adhm-abjm-bmn-triangle}
\end{figure}

The monopole charges and their multiplicities specify the BMN vacuum,
with the total topological charge giving the BMN matrix size.
For $N_f=1$, this charge is momentum along the M-theory circle, and
ADHM theory is expected to share an infrared fixed point with level-one
ABJM theory
\cite{Hayashi:2022ldo},
\begin{equation}
 U(N)\ \text{ADHM with one fundamental}
 \quad\longleftrightarrow\quad
 U(N)_1\times U(N)_{-1}\ \text{ABJM}.
 \label{eq:adhm-abjm-duality}
\end{equation}
Together with the ABJM--BMN index relation of
Ref.~\cite{Chang:2026yeg}, our result motivates the comparison in
Fig.~\ref{fig:adhm-abjm-bmn-triangle}. An explicit operator map should
identify the protected excitations in each ADHM monopole sector with
gauge-invariant ABJM operators built from monopoles and bifundamental
fields, matching their refined charges.

\begin{figure}[htbp]
 \centering
 \begin{tikzpicture}[
   theory/.style={align=center,inner sep=6pt,font=\small},
   relation/.style={<->,>=Latex,thick},
   relabel/.style={fill=white,align=center,font=\scriptsize,inner sep=2pt}]
  \node[theory] (adhm) at (-4.5,1.4) {$N_f=1$ ADHM theory};
  \node[theory] (other) at (4.5,1.4) {$\mathcal N=8$ SYM};
  \node[theory] (bmn) at (0,-2) {BMN matrix model};
  \draw[relation] (adhm.east) -- (other.west);
  \node[relabel,text width=3.6cm] at (0,1.85) {mirror IR duality\\of full theories};
  \draw[relation,dashed] (adhm.south east) to[bend right=12] (other.south west);
  \node[relabel,text width=3.8cm] at (0,0.65) {identification of monopole sectors\\(open problem)};
  \draw[relation] (adhm.south) --
   node[relabel,pos=0.55,left=3mm,text width=2.6cm]
   {index matching at large charge\\(this work)} (bmn.west);
  \draw[relation,dashed] (other.south) --
   node[relabel,pos=0.55,right=3mm,text width=2.6cm]
   {proposed sector\\comparison} (bmn.east);
 \end{tikzpicture}
 \caption{For $N_f=1$, the proposed comparison with mirror SYM. The dashed
 edges indicate the proposed correspondence between fixed ADHM monopole
 sectors and SYM contributions, and the associated comparison with BMN.}
 \label{fig:adhm-sym-bmn-triangle}
\end{figure}
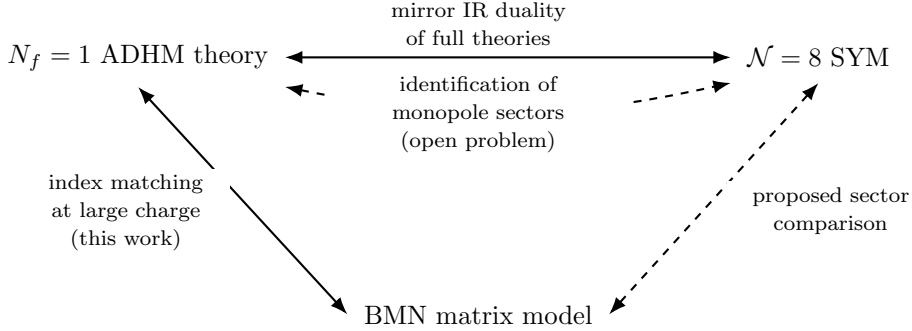

In the mirror $\mathcal N=8$ SYM description, ADHM topological charge
becomes flavor charge. Identifying the SYM contributions corresponding
to fixed ADHM monopole sectors would allow the comparison with BMN at
large flavor charge shown in Fig.~\ref{fig:adhm-sym-bmn-triangle}.

At finite charge, it remains to determine the leading corrections for
general vacuum partitions and identify cancellations that improve the
bounds. For $N_f=1$, identifying the ABJM or mirror-SYM operators
responsible for the ADHM fundamental corrections would further test
the proposed operator map.

It would be interesting to compare our large-charge limit with the triple scaling limit of \cite{Komatsu:2026ano, Komatsu:2026part2}.
In their construction, the ABJM rank, Chern--Simons level and angular
momentum along the Hopf fiber grow together, while the BMN matrix size
and coupling remain fixed.

\section*{Acknowledgments}
We thank Chi-Ming Chang for helpful discussions. Sarthak Duary is supported by the
Shuimu Tsinghua Scholar Program of Tsinghua University and the Beijing
Natural Science Foundation of China Grant No.~IS25035. Kangning Liu is supported
by the National Natural Science Foundation of China special fund for
theoretical physics No.~12447108 and the
national key research and development program of China
No.~2020YFA0713000. Kangning Liu also acknowledges support from the Tsinghua
University Short-term Visiting Scholarship for Doctoral Students.

\bibliographystyle{JHEP}
\bibliography{references}

\end{document}